\documentclass[amsmath,amssymb,prb]{revtex4}
\usepackage{graphicx}
\usepackage[cp1251]{inputenc}
\usepackage{amsmath}

\begin{document}

\title{Interaction of linear waves and moving Josephson vortex
lattices in layered superconductors}
 \author{A. V. Chiginev}
\email{chig@ipm.sci-nnov.ru}
\author{V. V. Kurin}
\affiliation{Institute for Physics of Microstructure of the
Russian Academy of Science, GSP-105, 603950 Nizhny Novgorod,
Russia}
\date{\today}

\begin{abstract}
A general phenomenological theory describing dynamics of Josephson vortices coupled to wide class of linear waves
in layered high-T$_c$ superconductors is developed. The theory is based on hydrodynamic long wave approximation and
describes interaction of vortices with electromagnetic, electronic and phonon degrees of freedom on an equal footing. In the limiting cases the proposed
theory degenerates to simple models considered earlier. In the framework
of the  suggested model we undertook the numerical simulation of resistive
state in layered superconductors placed in external magnetic field and demonstrate excitation of linear waves of various origin by a moving
vortex lattice, manifesting in existence of resonances on current-voltage characteristics.
\end{abstract}


 \maketitle

\section{Introduction}

High-T$_c$ superconductors (HTSCs) with strong anisotropy has been
the subject of intensive investigation during many years. The
layered structure, intrinsic Josephson effect, and complex
chemical composition provide a variety of physical properties of
such materials. A great deal of attention has been paid to
Josephson dynamics of layered superconductors.

The application of an external magnetic field parallel to the layers to a
layered HTSC leads to formation  of a Josephson vortex lattice (JVL) which can
be moved under the influence of an external current applied perpendicular to
the layers. When this lattice collides with the edge of the structure the
electromagnetic radiation is generated. This principle may be used for
construction of small and efficient electromagnetic oscillators with the
frequency being limited from above by the energy gap in HTSC which is of the
order of 10 THz. Recently, the radiation with the frequency of 0.85 THz and
power of 0.5 $\mu$W has been obtained from the structure based on
Bi$_2$Sr$_2$CaCu$_2$O$_8$~\cite{Ozyuzer_etal_Science_2007}. The small and
compact sources of electromagnetic radiation of terahertz band are very
interesting for applications in astrophysics, medicine, biology and many other
branches of science. Therefore, the investigation of Josephson dynamics in
layered HTSCs is highly important.

Layered HTSCs represent complex physical system with wide spectrum of
eigenwaves, which includes electromagnetic, plasma, different phonon modes and
Carlson-Goldman mode~\cite{Karlson_Goldman_PRL_1975}. All these modes are
coupled to vortices and can be excited by the moving JVL. The reverse action of
the excited waves on JVL resulting in changes of vortex shape and mutual
arrangement of vortices will affect the current-voltage characteristics (CVCs)
of layered superconductor and intensity of electromagnetic radiation. Thus
coupling of linear modes and moving vortices can be practically used for
controlling the dispersion properties of eigenwaves by the external magnetic
field, for diagnostics of eigenwaves by methods of Josephson
spectroscopy~\cite{Ponomarev_UFN_2002_eng}, and for controlling electromagnetic
radiation from layered superconductor.

The model adequately describing vortex dynamics in layered HTSCs should take
into account the interaction of vortices with all weakly damping linear modes.
However, the development of the theory of such comprehensive kind is not a
simple task and a main obstacle is a nonlocality of the constitutive equations
of a layered superconductor or, in other words, spatial dispersion. Even in
hydrodynamic approximation when the dynamical equations are differential, the
consistent approach to description of layered media requires solution of
differential equations for field and matter in the domains of homogeneity, and
subsequent joining of the solutions found for adjacent domains on interfaces.
If the number of modes in the system is rather high then the exact description
leads to sophisticated problem for eigenvectors and eigenvalues of
high-dimension transfer matrix. Example of calculations of this kind applied to
normal nonsuperconducting layered media can be found in
Ref.~\onlinecite{Mochan_etal_PRB_1987}. The description of layered systems can
be considerably simplified by employing the long-wavelength limit which deals
with smoothed  dynamical variables averaged over the spatial period of the
layered structure. The goal of the present paper is the development of such a
theory for a layered superconductor with intrinsic Josephson effect. We propose
an averaged, sufficiently simple, hydrodynamic theory accounting for a wide
spectrum of linear waves and vortex degrees of freedom. The model is easily
extendable for including additional linear modes and mechanisms of their
exitation.

By present, a number of models describing Josephson dynamics of layered HTSCs
in some special cases has been proposed. The first and most known one is the
local model accounting only for the magnetic (inductive) coupling between
adjacent junctions of the stack which was applied to description of the
artificial multilayer Josephson structures~\cite{Sakai_Bodin_Pedersen_JAP_1993}
and later to layered superconductors with intrinsic Josephson
effect~\cite{Bulaevskii_PRB_1994}. The charge coupling between the adjacent
junctions in layered superconductors has first been investigated in
Ref.~\onlinecite{Koyama_Tachiki_PRB_1996} for the case of spatially uniform
distributions of Josephson phase. There have been also some attempts to combine
magnetic and charge couplings into one
model~\cite{Kim_Pokharel_PhysicaC_2003,Machida_Sakai_PRB_2004}. The "global"
coupling of the junctions via the external waveguide connected parallel to the
long Josephson junction stack has been considered in our previous
paper~\cite{Chiginev_Kurin_PRB_2004}. The influence of nonequilibrium effects,
such as quasiparticle imbalance, to the Josephson dynamics of intrinsic
junctions in layered HTSCs, has been investigated in
Refs.~\onlinecite{Ryndyk_JETPLett_1997_eng,
Ryndyk_PRL_1998,Ryndyk_JETP_1999_eng,Ryndyk_etal_PRB_2001}. The effects of the
in-plane dissipation to the Josephson vortex motion in layered HTSCs has been
studied in
Refs.~\onlinecite{Koshelev_PRB_2000,Koshelev_Aranson_PRL_2000,Koshelev_Aranson_PRB_2001}.
The role of phonons in Josephon dynamics has been considered in
Refs.~\onlinecite{Helm_etal_PRL_1997,Helm_etal_PRB_2000} for the case of direct
excitation of phonons by the electric field, in
Refs.~\onlinecite{Maksimov_etal_SSC_1999,Ivanchenko_Medvedev_JETP_1971_eng} for
phonon excitation due to phonon-assisted tunneling. All these models may be
unified into one theory describing the linear waves and Josephson vortices in
layered HTSCs.

In the present paper we formulate the comprehensive theory accounting for
vortex interaction with linear waves of different physical nature which unifies
the models mentioned above. The prescription to design a theory of such kind is
the following. Instead of exact consideration of high-T$_c$ superconductor as a
layered medium we consider it as a continuous anisotropic medium which consists
of superconducting and normal electrons and of several sort of ions. To
describe the dynamics of such a media coupled to electromagnetic field we use
Maxwell equations jointly with two-liquid anisotropic hydrodynamic equations
for electrons and equations of adiabatic Born-Oppenheimer approximation for
ions. But this set of equations describe only linear properties of layered
superconductor in long-wave limit. To introduce vortex degrees of freedom to
the model and allow finite jump of Josephson phase difference on dielectric
layers we "discretize" the set of equations simultaneously restoring the
nonlinear term with sinusoidal dependence on Josephson phase difference in
interlayer current. In various particular cases our model is reduced to the
known models considered earlier. In the framework of the  suggested model we
undertook the numerical simulation of resistive state in layered
superconductors placed in external magnetic field and demonstrate excitation of
linear waves of various origin by a moving vortex lattice, manifesting in
existence of resonances on current-voltage characteristics.

The paper is organized as follows. The next section is devoted to formulation
of effective averaged model describing linear waves and vortex degrees of
freedom in layered superconductors with intrinsic Josephson effect and to
estimation of effective parameters of the model. The third section is devoted
to numerical experiment showing the effects of excitation of linear waves by
Josephson vortex lattice moving in a layered HTSC. In the conclusion we
summarize the main results obtained in the paper.

\section{The phenomenological description of Josephson dynamics of a layered HTSC --- hydrodynamic approach.}

In this section we derive the set of equations describing Josephson dynamics of
a layered superconductor treating it as a continuous but anisotropic medium.
First, we write down the Maxwell equations and  anisotropic hydrodynamic
equations describing superconducting condensate, electronic quasiparticles and
phonon degrees of freedom of superconductor. As a result of this stage we
obtain a system describing linear waves in a layered HTSC in a long wavelength
limit. Then, substituting the derivatives over the coordinate perpendicular to
the layers by finite differences and, simultaneously,
 restoring the nonlinear expression for the interlayer supercurrent in Josephson form $j_s = j_c \sin\theta_n$, we obtain the desired set of equations.

\begin{figure}[ht]
\includegraphics[width=8cm]{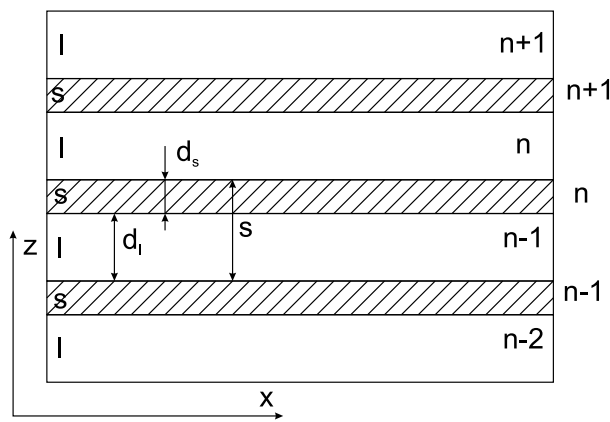}
\caption{Side view of the multilayer Josephson structure. The
numeration of superconducting and insulator layers, and the
coordinate system, are shown.}
\label{sketch}
\end{figure}

In this paper we adopt the widely used concept that high-T$_c$ superconductor
can be treated as a sequence of superconducting layers coupled through
dielectric layers by tunnel effect providing the existence of Josephson effect
and quasiparticle conductivity in $c$-axis direction. Such  a
layered superconductor  with chosen coordinate system is shown in
Fig.~\ref{sketch}. The $x$ axis is chosen laying in $ab$ plane of a
superconductor which is assumed to be isotropic in this plane, direction of the $y$ axis is chosen coinciding with the direction of the magnetic field which is assumed to have only one component, both static and alternating, and the $z$ axis is directed perpendicular to the layers along $c$ axis of layered superconductor. For such magnetic field the electric field and other physical variables would not depend on $y$ and the problem becomes two-dimensional one. The superconducting and adjacent dielectric layers are enumerated by number $n$, as shown in Fig.~\ref{sketch}.

We start the derivation from the assumption that, first, the characteristic
scale of the solutions is much greater than the layer structure period, and,
second, the module of the Josephson phase difference over dielectric layer is
much smaller than unity. These assumptions allow us to treat layered
superconductor as a continuous medium, and, thus, to formulate the desired
model in the continual limit. Then, after "discretizing" the continuous model,
we will restore possibility of finite jump of the phase of the order parameter and
finally obtain the set of finite-difference equations representing the desired
model allowing vortex solutions.

As the basic equations we use Maxwell equations for electromagnetic field together
with hydrodynamic equations describing contributions from superconducting electrons, normal
electrons, and phonons. The next subsections are devoted to derivation of
constitutive equations needed to construct the desired model.

\subsection{Field equations and superconducting electrons}

In this subsection we derive the set of equations phenomenologically describing
Josephson dynamics of a layered superconductor in a continual limit. To do
this, first we use Maxwell equations together with anisotropic hydrodynamic equations
for supercondicting electrons. Aftewards we supplement the set of obtained equations
with constitutive equations describing contributions
from normal electrons and phonons, which are to be derived in the next subsections.

As the starting point of the derivation of the equations describing the
superconducting electron subsystem we use the expression of the superconducting
momentum which we denote as $\mathbf p$ via the phase $\chi$ of superconducting order parameter
\begin{equation}
\mathbf p = \hbar \nabla \chi + \frac{2e}{c} \mathbf A ,
\label{gl}
\end{equation}
here  $\mathbf A$ is the vector potential of the electromagnetic field and the negative electron charge is explicitly taken into account. On this stage we suppose that there are no vortices in the material, so that $[\nabla \times \nabla\chi]
=0$. Accounting for this, we obtain
\begin{equation}
\mathbf B = \frac{c}{2e} [\nabla\times\mathbf p].
\end{equation}
Substituting this expression into Maxwell equation $[\nabla\times \mathbf B] = 4\pi c^{-1} \mathbf j_{tot} + c^{-1} \dot{\mathbf E}$ we get the evolutional equation for $\mathbf E$
\begin{equation}
\frac 1 c \dot{\mathbf E} = \frac c {2e} [\nabla\times[\nabla\times\mathbf p]] - \frac{4\pi}{c} \mathbf j_{tot}.\label{e_evol}
\end{equation}
The total current $\mathbf j_{tot}$ in~\eqref{e_evol} is the sum of the
supercurrent $\mathbf j_s$,  current of normal electrons $\mathbf j_n$, and ion
current $\mathbf j_i$. The supercurrent is related to velocity by usual
expression $\mathbf j_s =- e n_0 \mathbf v$ but in anisotropic Ginzburg-Landau
model the momentum is connected to velocity by relation $\mathbf p = 2\hat m_s
\mathbf v$ via tensor $\hat m_s$ of effective mass. We write down this tensor in
the form $\hat m_s = \mathrm{diag} (m/\Gamma_s , m\Gamma_s)$,
%\begin{equation}
%{\hat m}_s = \left(
%\begin{array}{cc}
%m/\Gamma_s & 0 \\ \\ 0 & m\Gamma_s
%\end{array}
%\right)
%\end{equation}
where  $m=(m_{xx} m_{zz})^{1/2}$ is the geometric average of the effective masses of superconducting
electrons in the in-plane and interlayer directions, factor $\Gamma_s=(m_{xx}/m_{zz})^{1/2}$
is a measure of anisotropy of superconducting properties of the material. Using the expression for supercurrent,
one rewrites~\eqref{e_evol} in components
\begin{subequations}
\begin{eqnarray}
\frac{1} c \dot E_x = \frac{c}{2e}\left(\frac{\partial^2 p_z}{\partial x \partial z} -\frac{\partial^2 p_x}{\partial z^2}\right) -
\frac{4\pi} c \left(-\frac{en_0\Gamma_s}{2m} p_x +{j_n}_x+{j_i}_x\right), \label{ex_evol} \\
\frac{1} c \dot E_z = \frac{c}{2e}\left(-\frac{\partial^2 p_z}{\partial x^2} +\frac{\partial^2 p_x}{\partial x \partial z}\right) -
\frac{4\pi} c \left(-\frac{en_0}{2m\Gamma_s} p_z +{j_n}_z+{j_i}_z - j_{ext}\right) .\label{ez_evol}
\end{eqnarray}
\end{subequations}

Though we assume that nothing depends on $y$-coordinate in our system, in the
Eq.~\eqref{ez_evol} we take into account that the $z$-component of
$[\nabla \times {\mathbf B}]$ actually consists of two terms $$ [\nabla \times {\mathbf B}] = \frac{\partial B_y}{\partial x} - \frac{\partial B_x}{\partial y} \equiv
\frac{\partial B_y}{\partial x} + \frac{4\pi} c j_{ext}; $$
with the last term
in the r. h. s. of this equation being proportional to the bias current
$j_{ext}$. The Eqs.~\eqref{ex_evol}, \eqref{ez_evol} are to be supplemented
with evolutional equations for $\mathbf{p}$.

If we differentiate Eq.~\eqref{gl} with respect to time and introduce chemical
potential of superconducting electrons $2\mu = \hbar \dot\chi -2e\varphi$ then
we come to equation of hydrodynamic type
\begin{equation}
\hat{m}\dot{\mathbf v} +\nabla \mu=-e\mathbf E .\label{p_evol}
\end{equation}
Further we make a model assumption that chemical potential of superconducting
electrons is defined by formula of degenerated Fermi gas $\mu=\varepsilon_F$ where
$\varepsilon_F=\hbar^2(3\pi^2 n)^{2/3}/2m$ is the Fermi energy and $n$ is a
concentration of superconducting electrons. Assuming that deviation of the
concentration $n_s$ from its equilibrium value $n_{0}$ is small we can
linearize equation ~\eqref{p_evol} what yields
\begin{equation}
\hat m_s {\bf \dot v}_s + \frac{2}{3}\varepsilon_{F,s}\frac{n_s}{n_0} = -e{\bf
E},
\end{equation}
where we denote as $\varepsilon_{F,s}=\varepsilon_{F}(n=n_0)$ a Fermi energy
for unperturbed superconducting concentration. Further, excluding the
concentration deviation $n_s$ from hydrodynamic equations using Maxwell
equation which we write in the form $(\nabla\cdot\mathbf D) = 4\pi\rho_s \equiv
-4\pi e n_s$, we come to evolutional equation for $\mathbf p$
\begin{equation}
\frac 1 {2e} \dot{\mathbf p} = \mathbf E - r^2_d \nabla(\nabla\cdot \mathbf D ),
\end{equation}
or, in components
\begin{subequations}
\begin{eqnarray}
-\frac 1 {2e} \dot p_x = E_x - r^2_d \left(\frac{\partial^2 D_x}{\partial x^2} + \frac{\partial^2 D_z}{\partial x \partial z}\right), \label{px_evol}\\
-\frac 1 {2e} \dot p_z = E_z - r^2_d \left(\frac{\partial^2 D_x}{\partial x \partial z} + \frac{\partial^2 D_z}{\partial
z^2}\right)\label{pz_evol}.
\end{eqnarray}
\end{subequations}
Here $r_d =(2/3)^{1/2} v_F /\omega_p$ is a screening length of longitudinal
electric field, $v^2_F = 2\varepsilon_F/m$, and $\omega^2_p = 4\pi e^2 n_0
m^{-1}$ are averaged Fermi velocity and plasma frequency respectively which are
defined as $v^2_F=({v^2_F}_x {v^2_F}_z)^{1/2}$, $\omega^2_p=({\omega^2_p}_x
{\omega^2_p}_z)^{1/2}$. Vector of electric displacement field $\mathbf D$ is
defined so that it includes charges of normal electrons and ions as bound charges,
i.e.
\begin{equation}
\mathbf D = \mathbf E + 4\pi \mathbf P_n +4\pi\mathbf P_i\equiv (\hat I +
4\pi\hat\chi_n +4\pi\hat\chi_i)\mathbf E ,
\end{equation}
where $\mathbf P_n$ and $\mathbf P_i$ are the polarizations associated with
normal electrons and phonons, respectively, so that \mbox{$(\nabla \cdot
\mathbf P_{n,i}) = -\rho_{n,i}$}. Tensors $\hat\chi_n,\,\hat\chi_i$ are
normal electrons and phonons susceptibility respectively.

A set of equations Eqs.~\eqref{ex_evol}, \eqref{ez_evol}, \eqref{px_evol},
\eqref{pz_evol} describes linear dynamics of electromagnetic field and
superconducting condensate in layered superconductor in the long wave limit.
Other degrees of freedom as normal quasiparticles and phonons of different kind
come in the Eqs~\eqref{ex_evol}, \eqref{ez_evol} via extra current densities $\mathbf{j}_n$, $\mathbf{j}_i$ and in the Eqs.~\eqref{px_evol},
\eqref{pz_evol} via diplacement vector $\mathbf D$. In the next two
subsection we find contributions from quasiparticles and phonons to
susceptibility and make the set of equations complete.

\subsection{Contribution from normal electrons}

To describe normal electrons we assume that they form degenerate Fermi gas and apply
hydrodynamical Thomas-Fermi approach  what gives the following equations
\begin{eqnarray}
\dot n_n + {n_0}_n (\nabla \cdot {\bf v}_n) =0, \nonumber \\
\hat m_n ({\bf \dot v}_n +\hat \nu {\bf v}_n)+\frac{2}{3}\frac
{{\varepsilon_F}}{{n_0}_n} \nabla n_n = e{\bf E}.
\label{hydro_n}
\end{eqnarray}
In the Eqs.~\eqref{hydro_n} ${\hat m}_n =\mathrm{diag}(m/\Gamma_n,m\Gamma_n)$
is the mass tensor of normal electrons, $\Gamma_n$ is the anisotropy parameter,
${n_0}_n$ is the unperturbed electron concentration, $n_n$ is the deviation of
electron concentration from its unperturbed value, ${\bf v}_n$ is the electron
velocity, ${\varepsilon_{F,n}}=\varepsilon_{F}(n=n_{0,n})$ is the Fermi energy
of normal electrons and tensor $ \hat\nu =\mathrm{diag}(\nu_x,\nu_z)$
characterizes the electron collision frequencies. For superconductor the
condition $\omega \ll \nu_{x,z}$ is usually satisfied and equations can be
somewhat simplified by neglecting the term ${\bf \dot v}$ in comparison with
$\hat \nu {\bf v}$. Then such hydrodynamic equations can be solved and the
relation between normal current ${ \bf j}_n$ and electric field ${\bf E}$ can be
easily obtained. In $(\omega,{\bf k})$-representation one gets
\begin{equation}
{\bf j}_n = \frac{{\omega^2_p}_n}{4\pi\Delta} \left(
\begin{array}{cc}
3i\omega\nu_z\Gamma_n - 2{v^2_F}_n k^2_z & 2{v^2_F}_n k_x k_z \\ \\
2{v^2_F}_n k_x k_z & 3i\omega\nu_x/\Gamma_n - 2{v^2_F}_n k^2_x
\end{array}
\right) {\bf E} \equiv -i\omega \hat\chi_n \mathbf E.\label{jqe}
\end{equation}
Here ${\omega^2_p}_n = 4\pi e^2 {n_0}_n /m $ is the plasma frequency, ${v_F}_n
= ({\varepsilon_F} /m)^{1/2}$ is the Fermi velocity, both related to normal
electrons, $\Delta=3i\omega \nu_x \nu_z -2\nu_x \Gamma^{-1}_n {v^2_F}_n k^2_z -
2\nu_z \Gamma_n {v^2_F}_n k^2_x$. Without spatial dispersion of normal
electrons, i. e. at ${v_F}_n = 0$, this equation became the simple anisotropic
Ohm law
\begin{equation}
{\bf j}_n = \hat\sigma {\bf E} , \quad \hat\sigma =\mathrm{diag}(\sigma_{xx},\sigma_{zz}), \label{normal1}
\end{equation}
where $\sigma_{xx} = {\omega^2_p}_n \Gamma_n / (4\pi \nu_x), \sigma_{zz} =
{\omega^2_p}_n  / (4\pi \nu_z\Gamma_n)$ are respectively the in-plane and
interlayer normal conductivities. Here it is worth to mention that such local
model of normal conductivity has been used in works
~\cite{Koshelev_PRB_2000,Koshelev_Aranson_PRL_2000,Koshelev_Aranson_PRB_2001}
devoted to study the influence of the in-plane normal conductivity to the
dynamics and stability of moving JVLs in layered HTSCs. Our approach naturally
introduces into consideration the spatial dispersion originated from nonzero
pressure of normal electrons.

\subsection{Phonon contribution}

This subsection is devoted to accounting for phonons in our phenomenological
model describing Josephson dynamics of layered HTSCs. In the present paper we
take into account only the direct excitation of infrared-active phonons by the
electric field of Josephson oscillations. In fact here we will actually
generalize the approach used in Refs.~\onlinecite{Helm_etal_PRL_1997,
Helm_etal_PRB_2000} to the case of non-uniform solutions in distributed
Josephson systems. The nonlinear interaction between Josephson oscillations and
phonons due to effects of phonon assisted tunneling considered in
Refs.~\onlinecite{Ivanchenko_Medvedev_JETP_1971_eng,Maksimov_etal_SSC_1999}
would not be considered in this subsection, but later, when we formulate general
nonlinear equation, we will show how this mechanism can
be introduced in our model.

In order to find the phonon contribution to the dielectric permittivity we use
standard approach (see, for example, Ref.\onlinecite{Ziman_Principles_SSPh_eng}). Let us write the classical equation
of motion for ions derived in Born-Oppenheimer adiabatic approximation,
\begin{equation}
M_\nu  {\ddot{\mathbf z}^{\,\nu}_{\mathbf N}} = -\sum_{\mu,
\mathbf M} \hat G^{\,\nu\mu}_{\mathbf N- \mathbf M} {\mathbf
z}^{\,\mu}_{\mathbf M} + q_\nu {\mathbf E}^\nu_{\mathbf N} .
\label{adiabatic}
\end{equation}
In this formula ${\mathbf z}^{\,\nu}_{\mathbf N}$ is the ion
displacement from the equilibrium position, $\nu$ is the ion index in the
unit cell, $\mathbf N =n_1 {\mathbf a}_1 +n_2 {\mathbf a}_2 + n_3
{\mathbf a}_3$ is the unit cell index, ${\mathbf a}_{1,2,3}$ are
lattice periods, $\hat G^{\,\nu\mu}_{\mathbf N- \mathbf M}$ is the
"seed" force tensor resulting from the interaction between ions
via valence electrons, $q_\nu$ is the ion charge, $M_\nu$ is the
ion mass, ${\mathbf E}^\nu_{\mathbf N}$ is the microscopic
electric field at the point of the ion. The interaction between
ions and conductivity electrons is carried out via the last term
in the r.~h.~s. of~(\ref{adiabatic}). In the
equation~(\ref{adiabatic}) and further the product of a tensor and a vector is written in components
as $(\hat A \mathbf x )_i = A_{ij} x_j$. The relation between
microscopic field and average macroscopic field can be written via
so called Lorentz tensor $\hat L^{\,\nu\mu}_{\mathbf N - \mathbf M}$
\begin{equation}
{\mathbf E}^\nu_{\mathbf N} ={\mathbf E}_{\mathbf
N}+4\pi\sum_{\mu, \mathbf M} \hat L^{\,\nu\mu}_{\mathbf N -\mathbf
M} {\mathbf P}^\mu_{\mathbf M},
\end{equation}
where ${\mathbf P}^\mu_{\mathbf M}=q_\mu {\mathbf
z}^{\,\mu}_{\mathbf M}$ is the dipole momentum of the $\mu$-th ion
in the $\mathbf M$-th cell. With the account for the difference
between microscopic electric field from macroscopic field the
equation for $\nu$-th ion motion takes the form
\begin{equation}
M_\nu   {\ddot{\mathbf z}^{\,\nu}_{\mathbf N}} = -\sum_{\mu,
\mathbf M} \hat F^{\,\nu\mu}_{\mathbf N -\mathbf M} {\mathbf
z}^{\,\mu}_{\mathbf M} + q_\nu {\mathbf E}_{\mathbf N},
\end{equation}
where $\hat F^{\,\nu\mu}_{\mathbf N -\mathbf M} = \hat
G^{\,\nu\mu}_{\mathbf N -\mathbf M} -4\pi q_\nu \hat
L^{\,\nu\mu}_{\mathbf N -\mathbf M} q_\mu$ is renormalized force tensor.

The contribution of ions into dielectric permittivity $\varepsilon$ may be found from the
expression for ion current density
\begin{equation}
{\mathbf j}_{\mathbf N} = \frac 1 V \sum_{\nu} q_\nu {\dot{\mathbf
z}^{\,\nu}_{\mathbf N}},
\end{equation}
where $V$ is the unit cell volume. After some algebra we get the
ion current density
\begin{equation}
\mathbf j_i = -\frac{i\omega} V \sum^{3L}_{a=1} \frac 1 {-\omega^2
+ \omega^2_{ph} (\mathbf k, a)} \frac{\sum_{\nu,\mu} q_\nu q_\mu
{\mathbf e}_\nu ( {\mathbf e}^{\,*}_\mu , \mathbf E )}{\sum_\nu
M_\nu {\mathbf e}^{\,*}_\nu {\mathbf e}_\nu} \equiv -i\omega \hat\chi_i\mathbf E, \label{phonon1}
\end{equation}
where $a$ is the phonon mode index, $L$ is the number of ions in
the unit cell, $\omega_{ph}(\mathbf k ,a)$ is the phonon
frequency, ${\mathbf e}_\nu (\mathbf k ,a)$ is the polarization
vector. $\omega_{ph}(\mathbf k ,a)$ and ${\mathbf e}_\nu (\mathbf
k ,a)$ are yielded from the equation for eigenvalues and
eigenvectors for the matrix $\hat F^{\nu\mu} (\mathbf k )$
\begin{equation}
\sum_\mu \hat F^{\nu\mu}(\mathbf k) \mathbf e_\mu (\mathbf k , a) = M_\nu \omega^2_{ph} (\mathbf k , a) \mathbf e_\nu (\mathbf k , a),
\end{equation}
where $\hat F^{\nu\mu} (\mathbf k )$ is the Fourier image  of the tensor $\hat F^{\,\nu\mu}_{\mathbf N -\mathbf M}$
\begin{equation}
\hat F^{\nu\mu} (\mathbf k ) = \sum_{\mathbf N}  e^{-i\mathbf
 k \mathbf N}\hat F^{\nu\mu}_{\mathbf N}.
\end{equation}
The expression~\eqref{phonon1} is the well-known formula describing phonon
contribution to dielectric permittivity, which may be found in many solid state
physics courses. Of course, the concrete definition of phonon frequencies and
polarization vectors for layered superconductors require detailed spectroscopic
or numerical investigation. Example of calculations of phonon properties of
typical layered superconductor with intrinsic Josephson effect Bi$_2$Sr$_2$CaCu$_2$O$_8$
can be found in Ref.~\onlinecite{Prade_etal_PRB_1989}.

So now we have found the contributions from normal electrons and phonons to the
dielectric permittivities and are able to define electric displacement vector $\mathbf D =   (\hat I +4\pi\hat\chi_n +
4\pi\hat\chi_i) \mathbf E \equiv \hat\varepsilon \mathbf E$, using
expressions~\eqref{jqe}
and~\eqref{phonon1} for susceptibilities $\hat\chi_n$ and $\hat\chi_i$.

The Eqs.~\eqref{ex_evol}, \eqref{ez_evol}, \eqref{px_evol}, \eqref{pz_evol}
together with the relations~\eqref{jqe} and~\eqref{phonon1} represent complete
system describing, in principle, all linear waves in long wavelength limit.
However, this system does not describe Josephson vortex degree of freedom in
layered HTSCs because it is linear and does not allow the finite jumps
of the phase of order parameter. In the next subsection we will show how
Josephson vortices can be included in the model and formulate general
nonlinear model describing both Josephson vortex dynamics and linear waves
in layered HTSCs. We will compare this model with several known ones, and reveal the
relations between the parameters of continual model and the parameters of
layered superconductors.

\subsection{Discretization of the model}

The aim of the present subsection is the transformation of the set of equations
in the continual limit, to the form admitting solutions in the form of
vortices, which are able to move along the layers of the HTSC. To do this, we
split the continious medium into the series of layers of thickness $s$ equal to
the period of the layers of the HTSC, and oriented perpendicular to the $c$
axis of superconductor. We introduce new variables describing the corresponding
fields at the moment $t$, coordinate $x$, and in some point within the $n$-th
layer. It is possible provided field distributions are smooth inside the
layer. After such discretization we allow the phase of order parameter to make
a finite jump between layers and restore the nonlinear expression for the
interlayer Josephson current.

First of all, let us decompose all dynamical variables to two sets defined on
two different lattices shifted with respect to each other. The nodes of these two lattices can be thought to be placed in the middle of superconducting and dielectric layers, respectively. The nodes of these sublattices we will denote by the number of period $n$. We attribute the variables $E_x,D_x,j_x^{n,s,i},v_x^{n,s,i}$
 to the first set and $E_z,D_z,j_z^{n,s,i},v_z^{n,s,i}$ to the
second one. After such a decomposition
we replace the derivatives of the dynamical variables over the
coordinate perpendicular to the layers in the set of Eqs.~\eqref{ex_evol},
\eqref{ez_evol}, \eqref{px_evol}, \eqref{pz_evol}, \eqref{jqe},
\eqref{phonon1}, by finite differences, using the following rule :
\begin{equation}
s\frac {\partial V}{\partial z} \to \pm V_{n\pm 1} \mp V_n ,
\qquad s^2 \frac{\partial^2 V}{\partial z^2} \to V_{n+1} -2V_n +
V_{n-1} \equiv \Delta_n V_n, \label{discretize}
\end{equation}
where $V_n$ stands for the one of the dynamical variables of the problem, upper
sign should be used for variables from the first set and lower sign --- for the
second one. Such a rule of discretisation follows from integral form of Maxwell and
hydrodynamic equations and provides necessary symmetry of finite-difference
equations. To demonstrate how this procedure works let us apply it to well known
telegraph equations describing electromagnetic waves in transmission lines
\begin{equation}
L\dot{I}+U_x=0,\quad C\dot{U}+I_x=0,
\end{equation}
here $U,I$ are voltage and current in the line, $L,C$ are linear densities of
inductance and capacity, respectively. Attributing voltage and current to different sets and replacing spatial derivatives by finite differences using our rule we come to equations
\begin{equation}
sL\dot{I}_n+U_n-U_{n-1}=0,\quad sC\dot{U}_n+I_{n+1}-I_n=0,
\end{equation}
expressing two Kirchhoff laws for discrete $L,C$ chain. Now let us return to
our problem. Applying the formulated procedure to our set of differential  the
substitution~\eqref{discretize} we obtain the following system
\begin{eqnarray}
\frac 1 c \frac{\partial }{\partial t}{E_x}_n = \frac{c}{2es} \frac{\partial}{\partial x} ({p_z}_n
- {p_z}_{n-1}) -\frac{c}{2es^2}
\Delta_n {p_x}_n  - \frac{4\pi} c
\left(-\frac{en_0\Gamma_s}{2m} {p_x}_n + {j^n_x}_n + {j^i_x}_n \right), \label{maxwell_x1} \\
\frac 1 c \frac{\partial }{\partial t}{E_z}_n = -\frac{c}{2e} \frac{\partial^2}{\partial x^2}
{p_z}_n - \frac{c}{2es}
\frac{\partial}{\partial x} ({p_x}_{n+1} - {p_x}_n) +
\frac{4\pi} c \left(-\frac{en_0}{2m\Gamma_s} {p_z}_n + {j^n_z}_n
+ {j^i_z}_n -j_{ext}\right), \label{maxwell_z1} \\
-\frac{1}{2e} \frac{\partial }{\partial
t}{p_x}_n = {E_x}_n - r^2_d
\left(\frac{\partial^2}{\partial x^2} {D_x}_n + \frac 1 s
\frac{\partial}{\partial x} ({D_z}_n - {D_z}_{n-1})\right), \label{koyama_x1} \\
-\frac{1}{2e}\frac{\partial }{\partial
t}{p_z}_n =  {E_z}_n -
r^2_d\left(\frac 1 s
\frac{\partial}{\partial x} ({D_x}_{n+1} - {D_x}_n) + \frac 1
{s^2} \Delta_n {D_z}_n \right), \label{koyama_z1} \\
\mathbf D = \hat\varepsilon \mathbf E,\quad \mathbf j^{n,i} = \frac\partial{\partial t}(\hat\chi_{n,i}\mathbf E), \quad \hat\varepsilon=\hat I +4\pi\hat\chi_n + 4\pi\hat\chi_i . \label{constitutive2}
\end{eqnarray}
The Eq.~\eqref{constitutive2} of this system combines the contributions from normal electrons and phonons which are given by the formulas~\eqref{jqe} and~\eqref{phonon1} for corresponding tensors of susceptibilities. In a common case, when spacial dispersion takes place, these tensors contain wavevectors in the interlayer direction. Therefore, the procedure of discretization should be applied also to Eq.~\eqref{constitutive2} accounting for specific expressions for normal electron and phonon contributions to dielectric permittivity.

Now this system allows finite jumps of physical variables but it still remains
linear and, therefore, does not describe Josephson vortices. To describe them we need to
introduce Josephson nonlinearity in the expression for interlayer supercurrent.
This procedure is made in the following way.
Integrating the expression for $z$-component of the superconducting momentum
over $z$ from $n$-th to $n+1$-th layer, we get Josephson phase difference
\begin{equation}
\theta_n = -\frac 1 \hbar \int^{(n+1)s+\frac{d_s} 2}_{ns+\frac{d_s} 2} p_z dz = \chi_n - \chi_{n+1}- \frac{2\pi}{\Phi_0} \int^{(n+1)s+\frac{d_s} 2}_{ns+\frac{d_s} 2} A_z dz .
\end{equation}
Then we need to substitute $\theta_n$ into the set of equations in a correct
way. As the interlayer supercurrent in layered HTSCs has Josephson nature, we
need to substitute ${p_z}_n \to -\hbar s^{-1}\sin\theta_n$ in the last term of
the Eq.~\eqref{maxwell_z1}. The remaining terms in Eqs.~\eqref{maxwell_x1}
and~\eqref{maxwell_z1} having ${p_z}_n$ are in fact the components of $[\nabla, \mathbf
B]$, that is why in these terms ${p_z}_n \to -\hbar s^{-1} \theta_n$. The
Eq.~\eqref{koyama_z1} at $r_d = 0$ is actually the Josephson relation $\hbar
\dot \theta = 2eU$, then in this expression also ${p_z}_n \to -\hbar s^{-1}
\theta_n$. At $r_d \neq 0$ the Eq.~\eqref{koyama_z1} describes violation of the Josephson relation due to charge effects; the influence of these effects on Josephson dynamics is considered later, in the section devoted to numerical experiment.

Let us rewrite the system~\eqref{maxwell_x1}, \eqref{maxwell_z1}, \eqref{koyama_x1}, \eqref{koyama_z1}, \eqref{constitutive2} in the notations used when describing bulk superconductors:
$$
j_c = \frac{\Phi_0}{8\pi^2 cs}\frac{{\omega^2_p}_s}{\Gamma_s}, \quad
\lambda^2_{ab} = \frac{c\Phi_0}{8\pi^2 j_c s \Gamma^2_s} =
\frac{c^2}{\Gamma_s {\omega^2_p}_s}, \quad r^2_d =
\frac{{v^2_F}_s}{{\omega^2_p}_s}\frac{d_s} s .
$$
As a result, we get the following set of equations:
\begin{eqnarray}
\frac 1 c \frac{\partial }{\partial t}{E_x}_n =
-\frac{\Phi_0}{2\pi s^2} \frac{\partial}{\partial x} (\theta_n -
\theta_{n-1}) +\frac{4\pi} c \frac{\lambda^2_{ab}}{s^2} \Delta_n
{j^s_x}_n - \frac{4\pi} c
({j^s_x}_n + {j^n_x}_n + {j^i_x}_n),\label{maxwell_x} \\
\frac 1 c \frac{\partial }{\partial t}{E_z}_n = \frac{\Phi_0}{2\pi
s} \frac{\partial^2}{\partial x^2} \theta_n - \frac{4\pi
\lambda^2_{ab} }{c s} \frac{\partial}{\partial x}({j^s_x}_{n+1} -
{j^s_x}_n) - \frac{4\pi} c (j_c \sin\theta_n + {j^n_z}_n
+ {j^i_z}_n - j_{ext}),\label{maxwell_z} \\
\frac{4\pi \lambda^2_{ab}}{c^2} \frac{\partial }{\partial
t}{j^s_x}_n = {E_x}_n - \frac{s r^2_d}{d_s}
\left(\frac{\partial^2}{\partial x^2} {D_x}_n + \frac 1 s
\frac{\partial}{\partial x} ({D_z}_n - {D_z}_{n-1})\right),\label{koyama_x} \\
\frac{\Phi_0}{2\pi c s}\frac{\partial }{\partial t}\theta_n =
{E_z}_n - \frac{s r^2_d}{d_s}\left(\frac 1 s
\frac{\partial}{\partial x} ({D_x}_{n+1} - {D_x}_n) + \frac 1
{s^2} \Delta_n {D_z}_n \right).\label{koyama_z} \\
\mathbf D = \hat\varepsilon \mathbf E,\quad \mathbf j^{n,i} = \frac\partial{\partial t}(\hat\chi_{n,i}\mathbf E), \quad \hat\varepsilon=\hat I +4\pi\hat\chi_n + 4\pi\hat\chi_i . \label{constitutive1}
\end{eqnarray}

The model represented by these equations ~\eqref{maxwell_x}, \eqref{maxwell_z},
\eqref{koyama_x}, \eqref{koyama_z}, \eqref{constitutive1} describes interaction
of Josephson vortices with electromagnetic waves, plasmons, and phonons on an
equal footing. This model possesses the necessary symmetry with respect to
permutation of $x,z$ coordinates and describes the spatial dispersion
resulted from electron and ion degree of freedom in the system. Besides, our system
allows to account for the influence of linear modes of any nature to the
dynamics of Josephson vortices in a layered HTSC, by inclusion of the
corresponding terms into the dielectric permittivity $\hat \varepsilon$. Say,
in recently discovered FeAs based superconductors\cite{Sadovskii_UFN_2008_eng, Ivanovskii_UFN_2008_eng, Izyumov_Kurmaev_UFN_2008_eng} it can
turn out that magnetic degrees of freedom play an important role. Our approach
allows to include them in this general scheme.

Of course, our general model contain the particular theories considered in
earlier works as a limiting cases. Among these special cases are the model with
the magnetic coupling between the
layers~\cite{Sakai_Bodin_Pedersen_JAP_1993,Bulaevskii_PRB_1994}, the model with
the charge coupling between the layers~\cite{Koyama_Tachiki_PRB_1996}, the
model taking into account the in-plane quasiparticle
current~\cite{Koshelev_PRB_2000}, the model accounting for the infrared-active
phonons polarized perpendicular to the
layers~\cite{Helm_etal_PRL_1997,Helm_etal_PRB_2000}, etc. Results of two works
\cite{Kim_Pokharel_PhysicaC_2003,Machida_Sakai_PRB_2004}, where the attempt to
unify magnetic and charge coupling have been undertaken also can be reproduced
by our model. But our model have extra terms in Eqs.~\eqref{koyama_x} and \eqref{koyama_z}
containing $\partial
/\partial x$ which are absent in Refs~\onlinecite{Kim_Pokharel_PhysicaC_2003,Machida_Sakai_PRB_2004}. Although small far from
resonances, these terms become significant when Josephson frequency get close
to phonon and plasmon frequencies.

The set of Eqs.~\eqref{maxwell_x}, \eqref{maxwell_z}, \eqref{koyama_x},
\eqref{koyama_z}, \eqref{constitutive1} contains the only nonlinear term, which
describes Josephson nonlinearity. While deriving this system we have neglected
all hydrodynamic nonlinearities such as nonlinear dependence of a pressure on
concentration, difference between Lagrange and Euler varyables, nonlinear
expression for the current density and so on. The account for these terms would
lead to the existence  of combination processes, such as, for exaple, Raman and
Brillouin  scattering. Due to strong anisotropy of the layered HTSCs, the most
significant effects of this type are the ones providing the dependence on
concentration of the parameters determining the transport and elastic
properties of the material in the direction of $c$-axis, such as Josephson
critical current density $j_c$, interlayer normal conductivity $\sigma_{zz}$,
etc. The accounting for these dependencies leads to such effects as
phonon-assisted tunneling due to the interaction of the electrons with
Raman-active phonons. This effect has been considered in
Ref.~\onlinecite{Ivanchenko_Medvedev_JETP_1971_eng} for acoustic phonons in
long Josephson junction and later in Ref.~\onlinecite{Maksimov_etal_SSC_1999}
for intrinsic junctions. The similar effect should be also expected for plasmons.
In the present work we restrict ourselves to the case when these effects are
unimportant.

Let us consider now the examples how some known models arise from our
equations in the limiting cases. For instance, if we set the phonon current,
the in-plane displacement and normal currents to be equal
to zero (${j_{x,z}}_{ph}=\partial {E_x}_n /\partial t = {j^n_x}_n = 0)$,
propose the interlayer normal current to have the purely Ohmic character
(${j_z}_n = \sigma_{zz} {E_z}_n$), and neglect the nonzero screening length of
the longitudinal  electric field $(r_d = 0)$, we obtain
\begin{equation}
\frac{\partial^2 \theta_n}{\partial x^2} = \left(\frac 1
{\lambda^2_j} \Delta_n + \frac 1 {\lambda^2_c}\right)
\left(\omega^2_j \frac{\partial^2 \theta_n}{\partial t^2} +
\sigma_{zz} \frac{\Phi_0}{2\pi c s} \frac{\partial \theta_n}{\partial
t} + \sin\theta_n -\frac{j_{ext}}{j_c}\right),
\end{equation}
where $\lambda_j = \Gamma_s s$,
$\lambda_c = \Gamma_s \lambda_{ab}$, $\omega^2_j = (8\pi^2 j_c
cs)/\Phi_0$. First this set of equations has been derived
in Refs.~\onlinecite{Sakai_Bodin_Pedersen_JAP_1993,Bulaevskii_PRB_1994} and describes the Josephson dynamics of stacked distributed Josephson junctions and layered superconductors with the magnetic coupling between the layers. In another case, when we suppose the lateral dimensions of the system to be smaller than Josephson length, so that $\partial / \partial x = 0$, neglect phonons, and set ${j_z}_n = \sigma_{zz}
{E_z}_n$, we obtain
\begin{equation}
\frac 1 {c^2} \frac{\partial^2\theta_n}{\partial t^2}  = -\frac 1
{\lambda^2_c} (1-\eta\Delta_n)\left(\sin\theta_n -\frac{j_{ext}}{j_c}\right)-
\frac{4\pi\sigma_{zz}}{c^2}\frac{\partial\theta_n}{\partial t},
\end{equation}
where $\eta = r^2_d/(sd_s)$ is the parameter of charge coupling. This set of equations has been first considered in Ref.~\onlinecite{Koyama_Tachiki_PRB_1996} and describes the dynamics of the layered superconductor with the charge coupling between the layers. Other particular examples of models considered earlier may be obtained from our model in a similar way.

\section{Dispersion characteristics of linear waves.}

The layered HTSCs has a wide spectrum of eigenwaves which, in principle, may be radiated by a moving JVL. The linear modes excited by JVL, in turn, may affect the shape and mutual arrangement of vortices in the lattice. The excitation of linear waves by the moving JVL leads to appearance of resonant steps on CVCs with the frequencies being equal to the ones of the radiated modes. In order to identify the resonances on CVC it is necessary to know the dispersion characteristics of linear waves in layered HTSCs, which can give much information about eigenmodes in the material and conditions of their excitation.

In order to build the dispersion characteristics we first linearize the set of Eqs.~\eqref{maxwell_x}, \eqref{maxwell_z}, \eqref{koyama_x}, \eqref{koyama_z}, \eqref{constitutive1} assuming $|\theta_n| \ll 1$, so that $\sin\theta_n \approx \theta_n$, and set the bias current to zero. We also neglect dissipation in the system ($\sigma_{xx}=\sigma_{zz}=0$) keeping in mind that it may be accounted for by perturbation theory. Taking all dynamic variables $\sim \exp{(ikx+iqn-i\omega t)}$ and setting the determinant of the obtained linear system to zero, we get the dispersion equation of the linear modes. It has the standard form
\begin{equation}
\det \left\| \frac{\omega^2}{c^2}{\varepsilon_{total}}_{ij}(\omega, \mathbf k) +k_i k_j -k^2 \delta_{ij} \right\| =0 ,
\end{equation}
where $\hat\varepsilon_{total}$ contains linear contributions from superconducting and normal electrons, and phonons. For short, we do not write this equation in an explicit form. We also assume the simple model expression of phonon susceptibility
\begin{equation}
\hat\chi_{ph}= - \frac 1 {4\pi}
\begin{pmatrix}
\frac{\displaystyle\Omega^2_x}{\displaystyle\omega^2+i\omega{\gamma_{ph}}_x-{\omega^2_0}_x} & 0 \\ 0 & \frac{\displaystyle\Omega^2_z}{\displaystyle\omega^2+i\omega{\gamma_{ph}}_z-{\omega^2_0}_z}
\end{pmatrix}. \label{phonon_susc}
\end{equation}
This expression describes two optical phonon modes: one polarized along the layers ($x$-phonon) and other polarized perpendicular to the layers ($z$-phonon). Here ${\omega_0}_x$ and ${\omega_0}_z$ are the frequencies of the $x$-phonon and the $z$-phonon, respectively, ${\gamma_{ph}}_x$ are ${\gamma_{ph}}_z$ the damping coefficients of these modes,
$\Omega_x$ and $\Omega_z$ are the oscillator strengths. For simplicity we assume ${\omega_0}_{x,z}$, ${\gamma_{ph}}_{x,z}$, and $\Omega_{x,z}$ to be independent of the quasimomentum, i. e. the bare phonon modes has no spatial dispersion. Moreover, in this section we neglect phonon damping (${\gamma_{ph}}_{x,z}=0$), assuming that it may be accounted for by perturbation theory. We will return to the nonzero ${\gamma_{ph}}_{x,z}$ later, in numerical experiment.

In our previous work we have considered wave vector surfaces for the analysis of the dispersion characteristics of the layered superconductor with phonons in the continual limit\cite{Chiginev_Kurin_SUST_2007}. Now we build the dispersion curves $\omega(k,q)$ for the periodic system.

\begin{figure}[h]
\centering
\includegraphics[width=8cm]{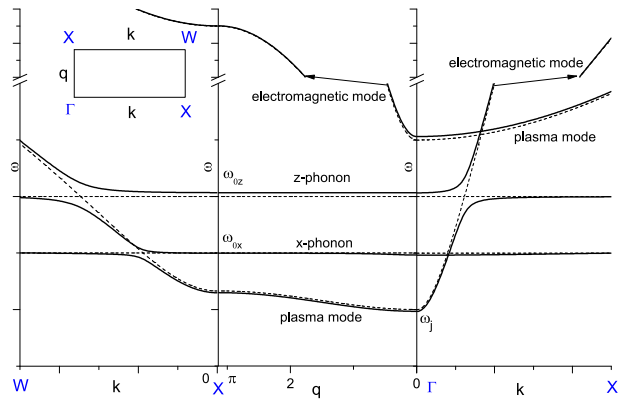}
\caption{(Color online) The schematic drawing of the dispersion characteristics of linear waves in a layered HTSC. The dashed lines show the dispersion without the interaction with phonon modes. The inset shows the fragment of the Brillouin zone of the layered structure.
}\label{dispersion}
\end{figure}

The dispersion curves of linear modes in a layered HTSC are schematically shown in Fig.~\ref{dispersion}. The electromagnetic, plasma, and phonon modes are shown. We build the dispersion characteristics for certain directions in the Brillouin zone. The curves in the right part of the plot are for the $\Gamma-{\rm X}$ direction (growing $k$), the curves in the central part of the plot are for the $\Gamma-{\rm X}$ direction (growing $q$), and the curves in the left part are for the ${\rm X}-{\rm W}$ direction (growing $k$), along the edge of the Brillouin zone. We use standard notations $\Gamma, {\rm X}, {\rm W}$ for the points of the Brillouin zone. The dispersion curves for the case of absence of the interaction between electromagnetic and plasma modes with phonons, i. e. at zero oscillator strengths ($\Omega_{x,z}=0$), are plotted by dashed lines. It is seen that at nonzero oscillator strengths the splitting of dispersion curves appears in the places of intersection between bare curves.

We note that due to anisotropy the separation of modes into the electromagnetic and plasma mode is relative, depending on the direction of the wave. For example, consider the lowest branch of the dispersion characteristics. The wave having $q=0$ is the electromagnetic mode and the one having $k=0$ is the plasma mode. And vice versa, for highest branch, the wave with $q=0$ is the plasma mode and the one with $k=0$ is the electromagnetic mode. For the arbitrary direction of the wave vector is it impossible to say whether it is an electromagnetic or plasma wave.

The frequency of the lowest branch of the dispersion characteristics corresponding to the point $\Gamma$ ($k=0, q=0$) is $\omega_j$---the frequency of the Josephson plasma resonance. In the notations of the hydrodynamic model it is equal to ${\omega_p}_s /\sqrt{\Gamma_s}$. The typical values of $\omega_j$ for layered HTSCs are of the order 100 GHz.
At the values of the parameters corresponding to layered HTSCs, the frequencies of the highest branch of the dispersion characteristics are in the optical range. Therefore, this mode cannot affect the JVL motion in the layered superconductors.

The important particular case of the modes in a layered HTSC is the Swihart wave, i. e. the mode belonging to the lowest branch of the dispersion characteristics, which propagating along the layers and having the standing wave structure in the direction perpendicular to the layers. The dispersion curve $\omega (k)$ of this wave has the form of the hyperbole. The slope of the asymptotes of this curve determines the Swihart velocity
\begin{equation}
\bar{c}^2(q) =\mu \frac{\varepsilon^{-1}_{zz}+2\eta(1-\cos q)}{1+2\mu(1-\cos q)}, \label{swihart}
\end{equation}
where $q$ is the transverse wave number of the Swihart wave.
It is seen that the antisymmetric ($q=\pi$) Swihart mode is the slowest one, and the symmetric ($q=0$) mode is the fastest one. We will use the formula~\eqref{swihart} for the analyzing the results of the numerical experiment.

\section{Numerical experiment}

Now we apply the derived system~\eqref{maxwell_x}, \eqref{maxwell_z},
\eqref{koyama_x}, \eqref{koyama_z}, \eqref{constitutive1} to the numerical investigation of the dynamics of Josephson vortices in layered superconductors with the account for their interaction with various linear modes.
The motion of Josephson vortex lattice (JVL) leads to excitation of linear modes of a layered HTSC which, generally, affect the moving lattice, leading to distortions in vortex shape and changes in their mutual arrangement. The excited linear modes may lead to resonances on CVCs of layered HTSCs with moving JVL. In the present section we numerically investigate the excitation of linear waves by the moving JVL, and their influence on CVCs, accounting for possible changes in vortex shape and their rearrangement.

In the numerical experiment we use periodic boundary conditions in both in-plane and interlayer directions for all variables. However, for Josephson phase difference $\theta_n$ the boundary conditions in $x$-direction is modified so that
\begin{equation}
\theta_n(L) = \theta_n (0) + 2\pi R_n ,
\end{equation}
where $R_n$ is the number of vortices trapped in the $n$-th junction of the stack, and $L$ is the length of the system in the longitudinal direction. Instead of $\theta_n$ we introduce $\theta'_n = \theta_n - 2\pi R_n x / L$ satisfying the periodic boundary condition in the longitudinal direction. Accounting for this, we write the expression for the interlayer Josephson current as $j_c \sin (\theta'_n +2\pi R_n x /L)$. The choice of boundary conditions described above provides the simplicity of numerical solution of the system; however, when using such conditions one takes into account only volume effects, neglecting the influence of boundaries on Josephson vortex dynamics.

In order to solve the system~\eqref{maxwell_x}, \eqref{maxwell_z},
\eqref{koyama_x}, \eqref{koyama_z}, \eqref{constitutive1} numerically we transform it to the set of evolutional equations in ordinary derivatives in time. To do this, we use the exponential Fourier transform by standard 2D-FFT algorithm. The obtained set of ordinary time-dependent differential equations is solved by Krank-Nikolson scheme. The similar approach to the numerical experiment has been used in Ref.~\onlinecite{Chiginev_Kurin_PRB_2004}.

The periodic boundary conditions along the layers imply that the number of vortices captured in each junction of the system is a constant. In our calculation we assume the number of vortices to be the same for all layers and denote it as $R$.

In the numerical experiment we use the typical values of the parameters of Bi$_2$Sr$_2$CaCu$_2$O$_8$: $j_c =
150$ A/cm$^2$, $\lambda_{ab} = 1700$ \AA, $s = 15$ \AA, $\sigma_{ab}
= 5 \cdot 10^4$ ($\Omega \cdot$ cm)$^{-1}$, $\sigma_c = 2 \cdot
10^{-3}$ ($\Omega \cdot$ cm)$^{-1}$, $\varepsilon_{zz}=12$. The parameters of phonons have been taken from the experimental data obtained by spectral ellipsometry\cite{Kovaleva_etal_PRB_2004}. In the following we assume the length of the system $L$ to be measured in the units of Josephson length $\lambda_j = \Gamma_s s$ and the bias current density to be measured in the units of Josephson critical current $j_c$. In the calculations we use values of $L$ corresponding to the ones of the patterns used in experiments. The number of vortices in a layer is set so that it corresponds to the dense JVL (at a given $L$). We also assume the frequencies and the voltages to be measured in the units of ${\omega_p} \Gamma^{-1/2}_s$.

\subsection{Results}

This subsection is devoted to the results of the numerical modeling of the dynamics of the JVL in a layered HTSC performed basing on the derived model. First, we investigate the simplest case, when the spatial dispersion, phonons, and all interlayer couplings except the magnetic one are neglected. Then, we complicate the problem step-by-step, introducing phonon susceptibility, charge coupling, and study the effects caused by these complications.

\subsubsection{The simplest case---the magnetic coupling without phonons and spatial dispersion.}

To check the efficiency of the used numerical scheme we perform the calculation of the CVC of the layered HTSC with the moving JVL in the case of absence of spatial dispersion, phonons, and in-plane dissipation. The interlayer normal current is assumed to be purely ohmic. This approximation corresponds to the case of magnetic coupling between the layers, which has been considered in Refs.~\onlinecite{Sakai_Bodin_Pedersen_JAP_1993,Bulaevskii_PRB_1994}.

\begin{figure}[ht]
\includegraphics[width=8cm]{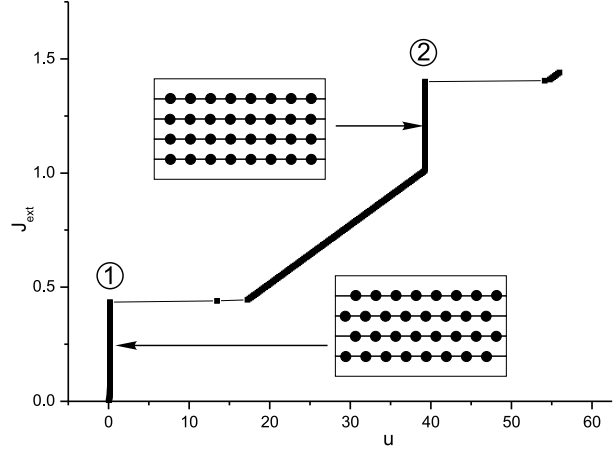}
\caption{A CVC of the layered HTSC with the moving JVL. The digits enumerate the steps of the CVC in the order of resonance frequencies increase. The insets schematically show the mutual arangement of vortices on different resonance steps.}
\label{vach2}
\end{figure}

\begin{figure}[ht]
\includegraphics[width=8cm]{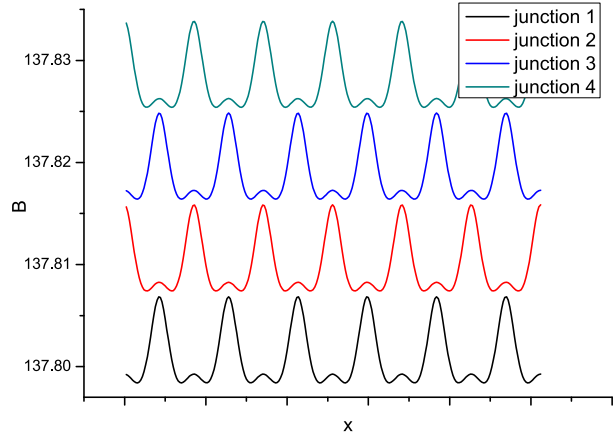}
\caption{(Color online) The magnetic field distribution
corresponding to the static vortex lattice ($j_{ext}=0$). For the sake of
convenience, here and in the subsequent figures we add constants to the distributions of the magnetic field in different junctions. The actual value of the constant component of the magnetic field is as for the junction 1.}
\label{j_0}
\end{figure}

Consider the CVC built for the case $R=6$ which corresponds to typical values of the external magnetic field (Fig.~\ref{vach2}). At $j_{ext}=0$ a triangular JVL is established in the structure (Fig.~\ref{j_0}). As the bias current increases, the JVL moves with the increasing velocity, keeping the triangular vortex arrangement. When the bias current reaches $0.18$, the JVL velocity stops growing, so that the voltage on the structure is established at the value $0.18$. This step denoted by 1 appears due to coincidence of the JVL velocity with the characteristic velocity of the antisymmetric $(q=\pi)$ Swihart mode of a layered HTSC.

The JVL remains triangular on the whole step 1; the vortex shape exhibit specific distortions due to the resonant increase of the harmonics with $q=\pi$ which grow as the bias current increases. The distribution of the magnetic field in this regime does not qualitatively differ from the static one (Fig.~\ref{j_0}).

\begin{figure}[ht]
\includegraphics[width=8cm]{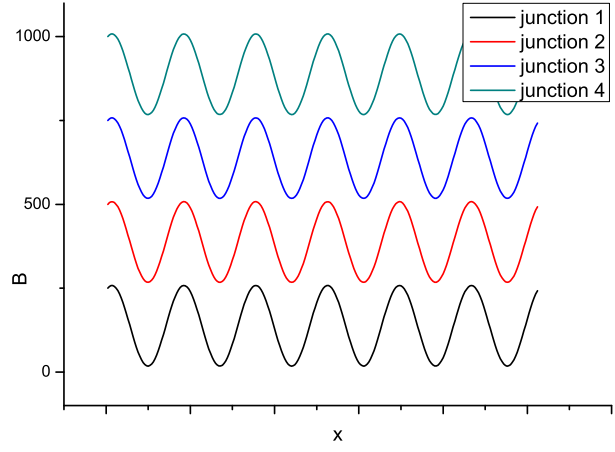}
\caption{(Color online) The magnetic field distribution
corresponding to the end of the step 2 in Fig.~\ref{vach2}. Here $j_{ext}=1.4$. Giant amplitude of the electromagnetic field.}
\label{j_14}
\end{figure}

At the value $j_{ext}= 0.44$ the CVC jumps from the step 1 to the Ohmic branch. The amplitude of the electromagnetic field on the Ohmic branch is small and the vortices form rectangular lattice. %(Fig.~\ref{j_06}).

When the bias current reaches the value $j_{ext}\approx 1.0$ the second step on the CVC appears. This step is due to the resonance with the symmetric ($q=0$) Swihart mode. In this regime, the moving vortices still form the rectangular lattice, but the amplitude of the electromagnetic field sharply increases.
This regime of vortex motion is characterized by large amplitude of the electromagnetic wave which accompanies the moving JVL, the amplitude of the wave grows with the bias current increase (Fig.~\ref{j_14}).

With further bias current increase the CVC again jumps to the Ohmic branch. The magnetic field distribution in this regime does not qualitatively differ from the one between the steps 1 and 2.

The CVC obtained in our calculations is similar to the one obtained for two-stacked long Josephson junctions~\cite{Petraglia_etal_JAP_1995}. The only difference is that in our CVC the voltages of the steps differ by two orders from each other. This is due to the fact that the magnetic coupling in layered HTSCs is usually much stronger than the one of artificial multilayer structures.

\begin{figure}[ht]
\includegraphics[width=8cm]{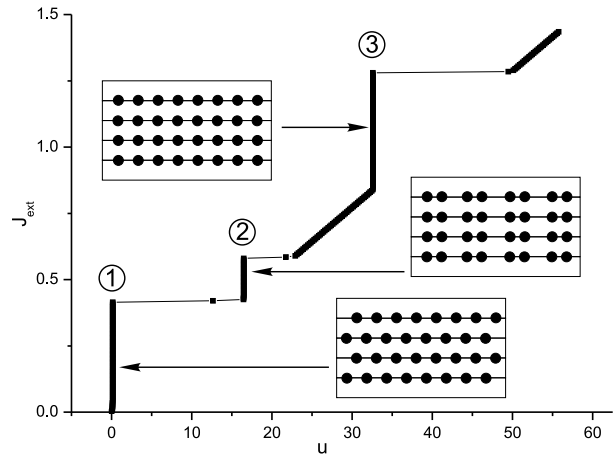}
\caption{CVC of the layered HTSC with moving JVL for weaker external magnetic field.}
\label{vach3}
\end{figure}

\begin{figure}[ht]
\includegraphics[width=8cm]{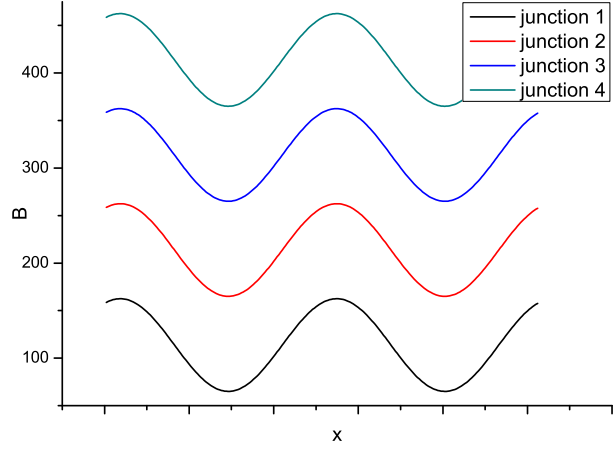}
\caption{(Color online) The magnetic field distribution
corresponding to the end of the step 2 in Fig.~\ref{vach3}. Here $j_{ext}=0.58$.}
\label{j_058}
\end{figure}

Consider now the CVC of the layered HTSC with the moving JVL at smaller external magnetic field ($R=4$) (Fig.~\ref{vach3}). One can see that, in addition to steps 1 and 3 corresponding to the resonances with the antisymmetric Swihart mode ($u=0.14$) and the symmetric mode ($u=32$), respectively, there is a step 2 corresponding to half a frequency of the resonance with the symmetric Swihart mode. The magnetic field distribution on this step is shown in Fig.~\ref{j_058}. It is seen that there are two oscillations of the magnetic field per spatial period of the system, though four magnetic flux quanta are captured in each junction of the stack. This regime also shows a large amplitude of the electromagnetic field. The distributions of the magnetic field and the mutual arrangement of vortices at the steps 1 and 3 and between the resonances are similar to the ones for the case considered earlier (see Figs.~\ref{vach2}, \ref{j_0}, \ref{j_14}).

\subsubsection{Excitation of a phonon by moving vortex lattice.}

The complex chemical composition of layered HTSCs provide large amount of phonon modes in such materials. Among these modes, there are "soft" phonons having frequencies of the order of several THz, which are smaller than the frequency of the energy gap in HTSCs. This makes possible the excitation of such phonon modes by a JVL moving in a layered HTSC. In this subsection we investigate the phonon excitation by a moving JVL.

For the calculations we slightly complicate the model used in the previous subsection, introducing the simplified expression~\eqref{phonon_susc} for the phonon susceptibility.
The phonon parameters used in calculations are taken from Ref.~\onlinecite{Kovaleva_etal_PRB_2004}.

\begin{figure}[h]
\centering
\includegraphics[width=8cm]{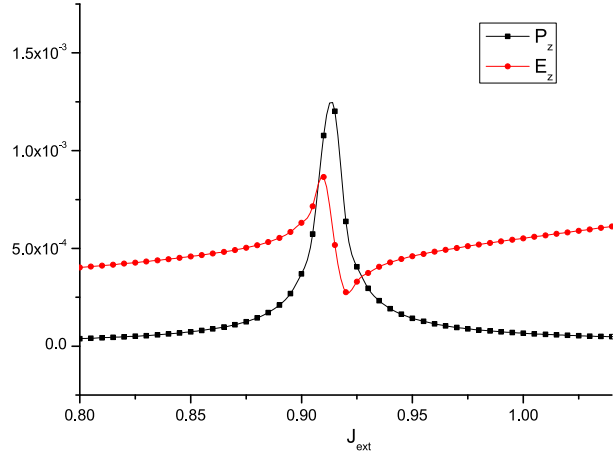}
\caption{(Color online) The amplitudes of alternate components of the electric
field and the polarization vs. external current in the vicinity of
the phonon frequency.
}\label{vach_phonon}
\end{figure}

\begin{figure}[h]
\centering
\includegraphics[width=8cm]{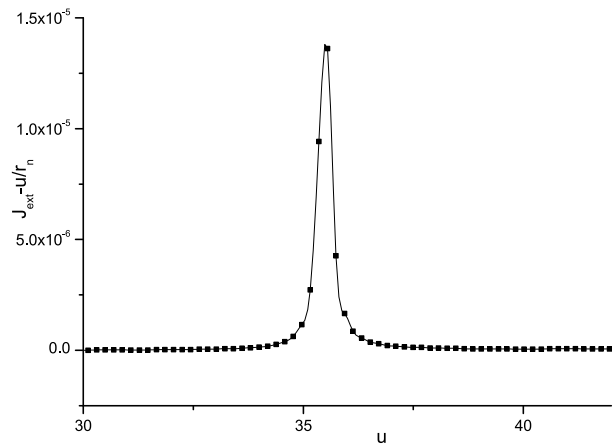}
\caption{Contribution to CVC due to excitation of the phonon mode in a HTSC.}\label{phonon_vach}
\end{figure}

The Figs.~\ref{vach_phonon} and~\ref{phonon_vach} illustrate the excitation of the phonon $z$-phonon by the moving JVL. The Fig.~\ref{vach_phonon} shows the dependence of complex amplitude modules of $P_z$ and $E_z$ harmonics with $q=0$ and $k=k_{lattice}$, on the bias current. Here $P_z$ is the $z$-component of the phonon polarization and $k_{lattice}=2\pi RL^{-1}$ is the wavenumber of the main harmonic of the JVL. The peak on the dependence $P_z (j_{ext})$ is due to the excitation of the phonon mode and appears at the value of $j_{ext}$ which corresponds to the JVL motion with the frequency of the phonon. The Fig.~\ref{phonon_vach} shows the peak on the contribution to the CVC due to the extra energy needed to excite the phonon. The similar phonon peaks on the CVC  has been obtained earlier~\cite{Helm_etal_PRL_1997,Helm_etal_PRB_2000} for the case of spatially uniform Josephson junctions and junction chains. Our model actually generalizes the one used in Ref.~\onlinecite{Helm_etal_PRB_2000} to the case of distributed Josephson juctions and junction stacks.

In the present simulations we do not consider the excitation of the $x$-phonons. The reason is that the JVL is rectangular at low enough external magnetic field and at the frequencies close to the phonon one. As one can see from the Eqs.~\eqref{maxwell_x}, \eqref{maxwell_z},
\eqref{koyama_x}, \eqref{koyama_z}, $x$-phonon is not excited by a rectangular JVL, as $E_x = 0$ in such a lattice. However, if the tensor of phonon susceptibility contains non-diagonal components, the excitation of $x$-phonons by $E_z$ of the moving  rectangular JVL is possible.

In this subsection we have considered the excitation of a phonon by a moving JVL provided the condition that the phonon frequency is far from the frequencies of Swihart modes at a given $k_{lattice}$. The next subsection is devoted to the situation when the phonon frequency coincides with the frequency of one of the Swihart modes.

\subsubsection{Excitation of the hybrid phonon+Swihart mode by a moving Josephson vortex lattice.}

\begin{figure}[h]
  \centering
\includegraphics[width=8cm]{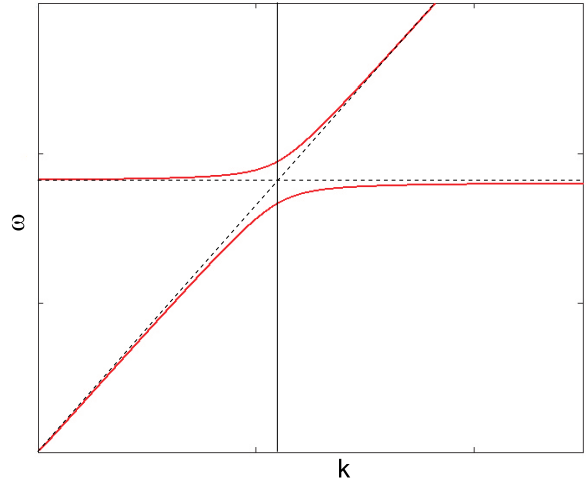}
\caption{(Color online) Dispersion characteristic of the symmetric Swihart mode and phonon mode in the vicinity of their interaction point.
}\label{dispersion1}
\end{figure}

\begin{figure}[h]
  \centering
\includegraphics[width=8cm]{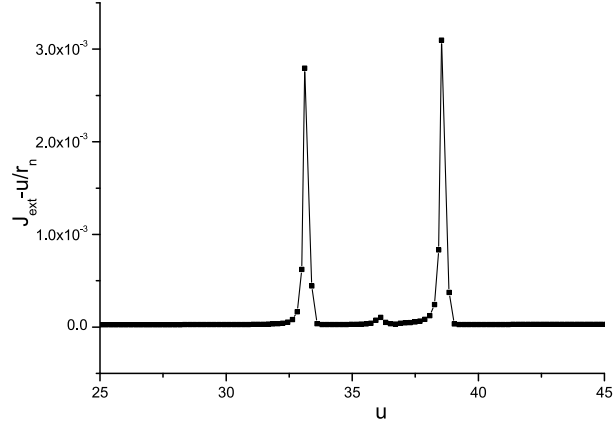}
\caption{Contribution to CVC due to the excitation of the hybrid phonon+Swihart modes.
}\label{mixed_mode}
\end{figure}

By choosing the external magnetic field applied to the structure, it is possible to make the frequency of the Swihart mode at $k=k_{lattice}$ equal to the phonon frequency. The Fig.~\ref{mixed_mode} shows the contribution to the CVC in the vicinity of the bias current value corresponding to the phonon frequency. Two peaks located close to each other and having nearly equal height appear due to excitation of two modes with close frequencies. To explain this effect, consider the fragment of the dispersion characteristic of linear waves in a layered HTSC in the vicinity of the point where the Swihart mode and the phonon mode interact (Fig.~\ref{dispersion1}). In the absence of the interaction the dispersion characteristic would have the shape shown by dotted line, here the slanted line shows the dispersion of the Swihart mode and the horizontal line shows the phonon dispersion.

In the presence of the interaction between two modes ($\Omega_x, \Omega_z \neq 0$) the dispersion characteristic takes the form shown by solid line in the Fig.~\ref{dispersion1}. The magnitude of the dispersion curve splitting is determined by the oscillator strength of the phonon mode ($\Omega_x$ or $\Omega_z$, depending on the polarization of the mode interacting with the Swihart wave). The intersection points of the vertical line corresponding to $k_{lattice}$, and dispersion curves, give the resonance frequencies.

The Fig.~\ref{mixed_mode} shows the contribution to the CVC from the excited hybrid phonon+Swihart modes. The distance between two peaks is equal to the difference between the resonance frequencies obtained from the dispersion characteristic. We note that the height of the peaks in Fig.~\ref{mixed_mode} is of two orders higher that the height of the peak caused by the excitation of the pure phonon mode (Fig.~\ref{phonon_vach}).

\subsubsection{The violation of the Josephson relation --- separation of normal electron charges by a moving JVL.}

In this subsection we investigate the influence of the charge effects on the JVL motion in layered HTSCs. Starting again from the simple model with the magnetic coupling, we now assume the parameter of the charge coupling to be nonzero $\eta \neq 0$ and find out the differences in JVL dynamics compared to the case $\eta=0$ considered above.

\begin{figure}[h]
  \centering
\includegraphics[width=8cm]{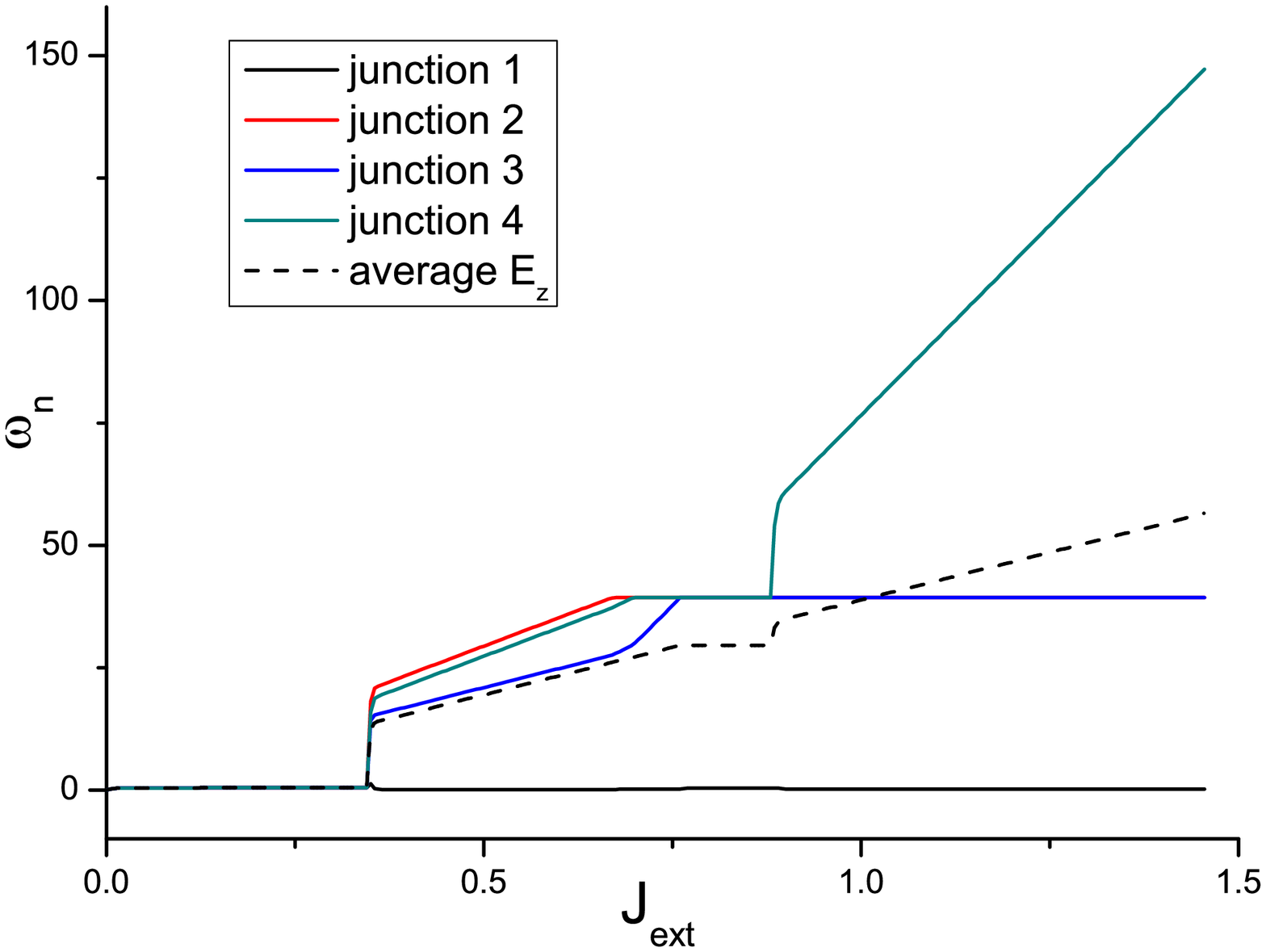}
\caption{(Color online) Dependence of phase growth rate in Josephson junctions of the structure, on the bias current. Dashed line shows the constant component of the electric field $E_z$ in each junction.
}\label{fach3}
\end{figure}

Consider CVC of the layered superconductor, calculated in the absence of phonons and in the presence of the charge coupling (Fig.~\ref{fach3}, dashed line). As it is seen from this figure, there are two steps of this CVC. As in the absence of the charge coupling, they appear due to resonance with the ancisymmetric and symmetric Swihart modes. Let us consider the differences between CVCs in Figs.~\ref{vach2} and \ref{fach3}. The first one is that the frequency of the first step is shifted towards higher frequencies, while the frequency of the second step is not changed. As in the case of zero charge coupling, the step positions correspond to the formula $\omega_{res} = \bar{c}(q)2\pi RL^{-1}$, where $\bar{c}(q)$ is determined by the expression~\eqref{swihart}.

The second feature of the regime with nonzero charge coupling is that the amplitude of the second step is much smaller than in the case of zero charge coupling. To explain this effect, consider the dependence of the Josephson phase growth rate on the bias current (Fig.~\ref{fach3}, solid lines). It is seen that the phase growth rates vary from one junction to another or, the same, the vortex chains in different junction have different velocities. At the same time, the voltages on each junction are the same. According to the Eq.~\eqref{koyama_z}, this is the demonstration of the violation of the Josephson relation. As it is also seen from Fig.~\ref{fach3}, only two or three junctions of possible four ones are locked to the second resonance of CVC. Therefore, the range of bias currents where the system remains on the resonant step, is smaller than in the case of zero charge coupling, when all junctions are locked.

\section{Conclusion}

We propose the comprehensive phenomenological model which describes the dynamics of the non-uniform distributions of Josephson phase difference in layered HTSCs, e. g. moving Josephson vortices and linear waves of
any nature. Basing on this system we numerically
build CVCs of a layered superconductor with the moving JVL and
demonstrate the excitation of linear modes by moving vortices. The proposed model is shown to cover many effects which have been studied in previous works; in addition, we observe some new effects such as excitation of a phonon and hybrid modes by a moving JVL in layered superconductors.

\section{Acknowledgements}

This work has been supported by the Russian Foundation for Basic
Research (Grant \# 09-02-01358-a), and by the following programs
of the Russian Academy of Science: "Nonlinear Dynamics", "Quantum
Macrophysics", and "Problems of Radiophysics".

\bibliography{literature}

\begin{thebibliography}{29}
\expandafter\ifx\csname natexlab\endcsname\relax\def\natexlab#1{#1}\fi
\expandafter\ifx\csname bibnamefont\endcsname\relax
  \def\bibnamefont#1{#1}\fi
\expandafter\ifx\csname bibfnamefont\endcsname\relax
  \def\bibfnamefont#1{#1}\fi
\expandafter\ifx\csname citenamefont\endcsname\relax
  \def\citenamefont#1{#1}\fi
\expandafter\ifx\csname url\endcsname\relax
  \def\url#1{\texttt{#1}}\fi
\expandafter\ifx\csname urlprefix\endcsname\relax\def\urlprefix{URL }\fi
\providecommand{\bibinfo}[2]{#2}
\providecommand{\eprint}[2][]{\url{#2}}

\bibitem[{\citenamefont{Ozyuzer et~al.}(2007)\citenamefont{Ozyuzer, Koshelev,
  Kurter, Gopalsami, Li, Tachiki, Kadowaki, Yamamoto, Minami, Yamaguchi
  et~al.}}]{Ozyuzer_etal_Science_2007}
\bibinfo{author}{\bibfnamefont{L.}~\bibnamefont{Ozyuzer}},
  \bibinfo{author}{\bibfnamefont{A.~E.} \bibnamefont{Koshelev}},
  \bibinfo{author}{\bibfnamefont{C.}~\bibnamefont{Kurter}},
  \bibinfo{author}{\bibfnamefont{N.}~\bibnamefont{Gopalsami}},
  \bibinfo{author}{\bibfnamefont{Q.}~\bibnamefont{Li}},
  \bibinfo{author}{\bibfnamefont{M.}~\bibnamefont{Tachiki}},
  \bibinfo{author}{\bibfnamefont{K.}~\bibnamefont{Kadowaki}},
  \bibinfo{author}{\bibfnamefont{T.}~\bibnamefont{Yamamoto}},
  \bibinfo{author}{\bibfnamefont{H.}~\bibnamefont{Minami}},
  \bibinfo{author}{\bibfnamefont{H.}~\bibnamefont{Yamaguchi}},
  \bibnamefont{et~al.}, \bibinfo{journal}{Science}
  \textbf{\bibinfo{volume}{318}}, \bibinfo{pages}{1291} (\bibinfo{year}{2007}).

\bibitem[{\citenamefont{Karlson and Goldman}(1975)}]{Karlson_Goldman_PRL_1975}
\bibinfo{author}{\bibfnamefont{R.~V.} \bibnamefont{Karlson}} \bibnamefont{and}
  \bibinfo{author}{\bibfnamefont{A.~M.} \bibnamefont{Goldman}},
  \bibinfo{journal}{Phys. Rev. Lett.} \textbf{\bibinfo{volume}{34}},
  \bibinfo{pages}{11} (\bibinfo{year}{1975}).

\bibitem[{\citenamefont{\mbox{Ya.}
  G.~Ponomarev}(2002)}]{Ponomarev_UFN_2002_eng}
\bibinfo{author}{\bibnamefont{\mbox{Ya.} G.~Ponomarev}},
  \bibinfo{journal}{Phys. Usp.} \textbf{\bibinfo{volume}{45}},
  \bibinfo{pages}{649} (\bibinfo{year}{2002}).

\bibitem[{\citenamefont{Moch\'an et~al.}(1987)\citenamefont{Moch\'an, del
  Castillo-Mussot, and Barrera}}]{Mochan_etal_PRB_1987}
\bibinfo{author}{\bibfnamefont{W.~L.} \bibnamefont{Moch\'an}},
  \bibinfo{author}{\bibfnamefont{M.}~\bibnamefont{del Castillo-Mussot}},
  \bibnamefont{and} \bibinfo{author}{\bibfnamefont{R.~G.}
  \bibnamefont{Barrera}}, \bibinfo{journal}{Phys. Rev. B}
  \textbf{\bibinfo{volume}{35}}, \bibinfo{pages}{1088} (\bibinfo{year}{1987}).

\bibitem[{\citenamefont{Sakai et~al.}(1993)\citenamefont{Sakai, Bodin, and
  Pedersen}}]{Sakai_Bodin_Pedersen_JAP_1993}
\bibinfo{author}{\bibfnamefont{S.}~\bibnamefont{Sakai}},
  \bibinfo{author}{\bibfnamefont{P.}~\bibnamefont{Bodin}}, \bibnamefont{and}
  \bibinfo{author}{\bibfnamefont{N.~F.} \bibnamefont{Pedersen}},
  \bibinfo{journal}{J. Appl. Phys.} \textbf{\bibinfo{volume}{73}},
  \bibinfo{pages}{2411} (\bibinfo{year}{1993}).

\bibitem[{\citenamefont{Bulaevskii et~al.}(1994)\citenamefont{Bulaevskii,
  Zamora, Baeriswyl, Beck, and Clem}}]{Bulaevskii_PRB_1994}
\bibinfo{author}{\bibfnamefont{L.~N.} \bibnamefont{Bulaevskii}},
  \bibinfo{author}{\bibfnamefont{M.}~\bibnamefont{Zamora}},
  \bibinfo{author}{\bibfnamefont{D.}~\bibnamefont{Baeriswyl}},
  \bibinfo{author}{\bibfnamefont{H.}~\bibnamefont{Beck}}, \bibnamefont{and}
  \bibinfo{author}{\bibfnamefont{J.~R.} \bibnamefont{Clem}},
  \bibinfo{journal}{Phys. Rev. B} \textbf{\bibinfo{volume}{50}},
  \bibinfo{pages}{12831} (\bibinfo{year}{1994}).

\bibitem[{\citenamefont{Koyama and Tachiki}(1996)}]{Koyama_Tachiki_PRB_1996}
\bibinfo{author}{\bibfnamefont{T.}~\bibnamefont{Koyama}} \bibnamefont{and}
  \bibinfo{author}{\bibfnamefont{M.}~\bibnamefont{Tachiki}},
  \bibinfo{journal}{Phys. Rev. B} \textbf{\bibinfo{volume}{54}},
  \bibinfo{pages}{16183} (\bibinfo{year}{1996}).

\bibitem[{\citenamefont{\mbox{Ju} H.~Kim and
  Pokharel}(2003)}]{Kim_Pokharel_PhysicaC_2003}
\bibinfo{author}{\bibnamefont{\mbox{Ju} H.~Kim}} \bibnamefont{and}
  \bibinfo{author}{\bibfnamefont{J.}~\bibnamefont{Pokharel}},
  \bibinfo{journal}{Physica C} \textbf{\bibinfo{volume}{384}},
  \bibinfo{pages}{425} (\bibinfo{year}{2003}).

\bibitem[{\citenamefont{Machida and Sakai}(2004)}]{Machida_Sakai_PRB_2004}
\bibinfo{author}{\bibfnamefont{M.}~\bibnamefont{Machida}} \bibnamefont{and}
  \bibinfo{author}{\bibfnamefont{S.}~\bibnamefont{Sakai}},
  \bibinfo{journal}{Phys. Rev. B} \textbf{\bibinfo{volume}{70}},
  \bibinfo{pages}{144520} (\bibinfo{year}{2004}).

\bibitem[{\citenamefont{Chiginev and Kurin}(2004)}]{Chiginev_Kurin_PRB_2004}
\bibinfo{author}{\bibfnamefont{A.~V.} \bibnamefont{Chiginev}} \bibnamefont{and}
  \bibinfo{author}{\bibfnamefont{V.~V.} \bibnamefont{Kurin}},
  \bibinfo{journal}{Phys. Rev. B} \textbf{\bibinfo{volume}{70}},
  \bibinfo{pages}{214523} (\bibinfo{year}{2004}).

\bibitem[{\citenamefont{Ryndyk}(1997)}]{Ryndyk_JETPLett_1997_eng}
\bibinfo{author}{\bibfnamefont{D.~A.} \bibnamefont{Ryndyk}},
  \bibinfo{journal}{JETP Lett.} \textbf{\bibinfo{volume}{65}},
  \bibinfo{pages}{791} (\bibinfo{year}{1997}).

\bibitem[{\citenamefont{Ryndyk}(1998)}]{Ryndyk_PRL_1998}
\bibinfo{author}{\bibfnamefont{D.~A.} \bibnamefont{Ryndyk}},
  \bibinfo{journal}{Phys. Rev. Lett.} \textbf{\bibinfo{volume}{80}},
  \bibinfo{pages}{3376} (\bibinfo{year}{1998}).

\bibitem[{\citenamefont{Ryndyk}(1999)}]{Ryndyk_JETP_1999_eng}
\bibinfo{author}{\bibfnamefont{D.~A.} \bibnamefont{Ryndyk}},
  \bibinfo{journal}{JETP} \textbf{\bibinfo{volume}{89}}, \bibinfo{pages}{975}
  (\bibinfo{year}{1999}).

\bibitem[{\citenamefont{Ryndyk et~al.}(2001)\citenamefont{Ryndyk, Pozdnjakova,
  Shereshevskii, and Vdovicheva}}]{Ryndyk_etal_PRB_2001}
\bibinfo{author}{\bibfnamefont{D.~A.} \bibnamefont{Ryndyk}},
  \bibinfo{author}{\bibfnamefont{V.~I.} \bibnamefont{Pozdnjakova}},
  \bibinfo{author}{\bibfnamefont{I.~A.} \bibnamefont{Shereshevskii}},
  \bibnamefont{and} \bibinfo{author}{\bibfnamefont{N.~K.}
  \bibnamefont{Vdovicheva}}, \bibinfo{journal}{Phys. Rev. B}
  \textbf{\bibinfo{volume}{64}}, \bibinfo{pages}{052508}
  (\bibinfo{year}{2001}).

\bibitem[{\citenamefont{Koshelev}(2000)}]{Koshelev_PRB_2000}
\bibinfo{author}{\bibfnamefont{A.~E.} \bibnamefont{Koshelev}},
  \bibinfo{journal}{Phys. Rev. B} \textbf{\bibinfo{volume}{62}},
  \bibinfo{pages}{R3616} (\bibinfo{year}{2000}).

\bibitem[{\citenamefont{Koshelev and
  Aranson}(2000)}]{Koshelev_Aranson_PRL_2000}
\bibinfo{author}{\bibfnamefont{A.~E.} \bibnamefont{Koshelev}} \bibnamefont{and}
  \bibinfo{author}{\bibfnamefont{I.~S.} \bibnamefont{Aranson}},
  \bibinfo{journal}{Phys. Rev. Lett.} \textbf{\bibinfo{volume}{85}},
  \bibinfo{pages}{3938} (\bibinfo{year}{2000}).

\bibitem[{\citenamefont{Koshelev and
  Aranson}(2001)}]{Koshelev_Aranson_PRB_2001}
\bibinfo{author}{\bibfnamefont{A.~E.} \bibnamefont{Koshelev}} \bibnamefont{and}
  \bibinfo{author}{\bibfnamefont{I.}~\bibnamefont{Aranson}},
  \bibinfo{journal}{Phys. Rev. B} \textbf{\bibinfo{volume}{64}},
  \bibinfo{pages}{174508} (\bibinfo{year}{2001}).

\bibitem[{\citenamefont{\mbox{Ch.} Helm et~al.}(1997)\citenamefont{\mbox{Ch.}
  Helm, \mbox{Ch.} Preis, Forsthofer, Keller, Schlenga, Kleiner, and
  \mbox{M\"{u}ller}}}]{Helm_etal_PRL_1997}
\bibinfo{author}{\bibnamefont{\mbox{Ch.} Helm}},
  \bibinfo{author}{\bibnamefont{\mbox{Ch.} Preis}},
  \bibinfo{author}{\bibfnamefont{F.}~\bibnamefont{Forsthofer}},
  \bibinfo{author}{\bibfnamefont{J.}~\bibnamefont{Keller}},
  \bibinfo{author}{\bibfnamefont{K.}~\bibnamefont{Schlenga}},
  \bibinfo{author}{\bibfnamefont{R.}~\bibnamefont{Kleiner}}, \bibnamefont{and}
  \bibinfo{author}{\bibfnamefont{P.}~\bibnamefont{\mbox{M\"{u}ller}}},
  \bibinfo{journal}{Phys. Rev. Lett.} \textbf{\bibinfo{volume}{79}},
  \bibinfo{pages}{737} (\bibinfo{year}{1997}).

\bibitem[{\citenamefont{\mbox{Ch.} Helm et~al.}(2000)\citenamefont{\mbox{Ch.}
  Helm, \mbox{Ch.} Preis, \mbox{Ch.} Walter, and Keller}}]{Helm_etal_PRB_2000}
\bibinfo{author}{\bibnamefont{\mbox{Ch.} Helm}},
  \bibinfo{author}{\bibnamefont{\mbox{Ch.} Preis}},
  \bibinfo{author}{\bibnamefont{\mbox{Ch.} Walter}}, \bibnamefont{and}
  \bibinfo{author}{\bibfnamefont{J.}~\bibnamefont{Keller}},
  \bibinfo{journal}{Phys. Rev. B} \textbf{\bibinfo{volume}{62}},
  \bibinfo{pages}{6002} (\bibinfo{year}{2000}).

\bibitem[{\citenamefont{Maksimov et~al.}(1999)\citenamefont{Maksimov, Arseyev,
  and Maslova}}]{Maksimov_etal_SSC_1999}
\bibinfo{author}{\bibfnamefont{E.~G.} \bibnamefont{Maksimov}},
  \bibinfo{author}{\bibfnamefont{P.~I.} \bibnamefont{Arseyev}},
  \bibnamefont{and} \bibinfo{author}{\bibfnamefont{N.~S.}
  \bibnamefont{Maslova}}, \bibinfo{journal}{Solid State Comm.}
  \textbf{\bibinfo{volume}{111}}, \bibinfo{pages}{391} (\bibinfo{year}{1999}).

\bibitem[{\citenamefont{\mbox{Yu.} M.~Ivanchenko and \mbox{Yu.}
  V.~Medvedev}(1971)}]{Ivanchenko_Medvedev_JETP_1971_eng}
\bibinfo{author}{\bibnamefont{\mbox{Yu.} M.~Ivanchenko}} \bibnamefont{and}
  \bibinfo{author}{\bibnamefont{\mbox{Yu.} V.~Medvedev}},
  \bibinfo{journal}{Sov. Phys.--JETP} \textbf{\bibinfo{volume}{33}},
  \bibinfo{pages}{1223} (\bibinfo{year}{1971}).

\bibitem[{\citenamefont{Ziman}(1972)}]{Ziman_Principles_SSPh_eng}
\bibinfo{author}{\bibfnamefont{J.~M.} \bibnamefont{Ziman}},
  \emph{\bibinfo{title}{Principles of the Theory of Solids}}
  (\bibinfo{publisher}{Cambridge at the University Press},
  \bibinfo{year}{1972}).

\bibitem[{\citenamefont{Prade et~al.}(1989)\citenamefont{Prade, Kulkarni,
  \mbox{de Wette}, Schroeder, and Kress}}]{Prade_etal_PRB_1989}
\bibinfo{author}{\bibfnamefont{J.}~\bibnamefont{Prade}},
  \bibinfo{author}{\bibfnamefont{A.~D.} \bibnamefont{Kulkarni}},
  \bibinfo{author}{\bibfnamefont{F.~W.} \bibnamefont{\mbox{de Wette}}},
  \bibinfo{author}{\bibfnamefont{U.}~\bibnamefont{Schroeder}},
  \bibnamefont{and} \bibinfo{author}{\bibfnamefont{W.}~\bibnamefont{Kress}},
  \bibinfo{journal}{Phys. Rev. B} \textbf{\bibinfo{volume}{39}},
  \bibinfo{pages}{2771} (\bibinfo{year}{1989}).

\bibitem[{\citenamefont{Sadovskii}(2008)}]{Sadovskii_UFN_2008_eng}
\bibinfo{author}{\bibfnamefont{M.~V.} \bibnamefont{Sadovskii}},
  \bibinfo{journal}{Phys. Usp.} \textbf{\bibinfo{volume}{51}},
  \bibinfo{pages}{1201} (\bibinfo{year}{2008}).

\bibitem[{\citenamefont{Ivanovskii}(2008)}]{Ivanovskii_UFN_2008_eng}
\bibinfo{author}{\bibfnamefont{A.~L.} \bibnamefont{Ivanovskii}},
  \bibinfo{journal}{Phys. Usp.} \textbf{\bibinfo{volume}{51}},
  \bibinfo{pages}{1229} (\bibinfo{year}{2008}).

\bibitem[{\citenamefont{\mbox{Yu.} A.~Izyumov and
  Kurmaev}(2008)}]{Izyumov_Kurmaev_UFN_2008_eng}
\bibinfo{author}{\bibnamefont{\mbox{Yu.} A.~Izyumov}} \bibnamefont{and}
  \bibinfo{author}{\bibfnamefont{E.~Z.} \bibnamefont{Kurmaev}},
  \bibinfo{journal}{Phys. Usp.} \textbf{\bibinfo{volume}{51}},
  \bibinfo{pages}{1261} (\bibinfo{year}{2008}).

\bibitem[{\citenamefont{Chiginev and Kurin}(2007)}]{Chiginev_Kurin_SUST_2007}
\bibinfo{author}{\bibfnamefont{A.~V.} \bibnamefont{Chiginev}} \bibnamefont{and}
  \bibinfo{author}{\bibfnamefont{V.~V.} \bibnamefont{Kurin}},
  \bibinfo{journal}{Supercond. Sci. Tech.} \textbf{\bibinfo{volume}{20}},
  \bibinfo{pages}{S34} (\bibinfo{year}{2007}).

\bibitem[{\citenamefont{Kovaleva et~al.}(2004)\citenamefont{Kovaleva, Boris,
  Holden, Ulrich, Liang, Lin, Bernhard, Tallon, Munzar, and
  Stoneham}}]{Kovaleva_etal_PRB_2004}
\bibinfo{author}{\bibfnamefont{N.~N.} \bibnamefont{Kovaleva}},
  \bibinfo{author}{\bibfnamefont{A.~V.} \bibnamefont{Boris}},
  \bibinfo{author}{\bibfnamefont{T.}~\bibnamefont{Holden}},
  \bibinfo{author}{\bibfnamefont{C.}~\bibnamefont{Ulrich}},
  \bibinfo{author}{\bibfnamefont{B.}~\bibnamefont{Liang}},
  \bibinfo{author}{\bibfnamefont{C.~T.} \bibnamefont{Lin}},
  \bibinfo{author}{\bibfnamefont{C.}~\bibnamefont{Bernhard}},
  \bibinfo{author}{\bibfnamefont{J.~L.} \bibnamefont{Tallon}},
  \bibinfo{author}{\bibfnamefont{D.}~\bibnamefont{Munzar}}, \bibnamefont{and}
  \bibinfo{author}{\bibfnamefont{A.~M.} \bibnamefont{Stoneham}},
  \bibinfo{journal}{Phys. Rev. B} \textbf{\bibinfo{volume}{69}},
  \bibinfo{pages}{054511} (\bibinfo{year}{2004}).

\bibitem[{\citenamefont{Petraglia et~al.}(1995)\citenamefont{Petraglia,
  Ustinov, Pedersen, and Sakai}}]{Petraglia_etal_JAP_1995}
\bibinfo{author}{\bibfnamefont{A.}~\bibnamefont{Petraglia}},
  \bibinfo{author}{\bibfnamefont{A.~V.} \bibnamefont{Ustinov}},
  \bibinfo{author}{\bibfnamefont{N.~F.} \bibnamefont{Pedersen}},
  \bibnamefont{and} \bibinfo{author}{\bibfnamefont{S.}~\bibnamefont{Sakai}},
  \bibinfo{journal}{J. Appl. Phys.} \textbf{\bibinfo{volume}{77}},
  \bibinfo{pages}{1171} (\bibinfo{year}{1995}).

\end{thebibliography}

\end{document}